\documentclass[11pt, tightenlines, onecolumn, amsmath, amssymb, notitlepage, a4paper, superscriptaddress]{revtex4-1}
\UseRawInputEncoding
\usepackage[strict]{revquantum}
\usepackage{graphicx}
\usepackage{algpseudocode} 
\usepackage{algorithm}
\usepackage{subfig}
\usepackage{braket}
\usepackage[margin=1in]{geometry}
\usepackage{bbm}
\usepackage{multirow}
\usepackage{xcolor}
\newtheorem{remark}[theorem]{Remark}

\DeclareMathOperator{\tr}{tr}

\algnewcommand\algin{\textbf{in}}

\begin{document}

\title{Multi-Axis Control of a Qubit in the Presence of Unknown Non-Markovian Quantum Noise}

\author{Akram Youssry}
\affiliation{Quantum Photonics Laboratory and Centre for Quantum Computation and Communication Technology, RMIT University, Melbourne, VIC 3000, Australia}
\author{Hendra I. Nurdin}
\affiliation{School of Electrical
Engineering and Telecommunications, UNSW Australia, Sydney, NSW 2052, Australia}

\begin{abstract}
In this paper, we consider the problem of open-loop control of a qubit that is coupled to an unknown fully quantum non-Markovian noise (either bosonic or fermionic). A graybox model that is empirically obtained from measurement data is employed to approximately represent the unknown quantum noise. The estimated model is then used to calculate the open-loop control pulses under constraints on the pulse amplitude and timing.
For the control pulse optimization, we explore the use of gradient descent and genetic optimization methods. We consider the effect of finite sampling on estimating expectation values of observables and  show results for single- and multi-axis control of a qubit. 

\end{abstract}

\keywords{quantum control, quantum noise, machine learning, open quantum systems}

\maketitle

\section{Introduction}
The importance of quantum control as an engineering task is increasing with the rapid development of quantum technologies, particularly quantum computing. The setting is a quantum system, such as a qubit, whose evolution can be steered by modulating some external controls. The objective is to find the optimal controls to achieve a desired target (such as a state or a gate). Many techniques have been developed for this purpose such as gradient-based methods \cite{khaneja2005optimal,caneva2011chopped,PhysRevA.95.042318,Haas_2019,PhysRevA.99.052327,Ciaramella_2015,de_Fouquieres_2011} where the fidelity between the target and the system is optimized with respect to the controls. The drawback of such methods is that they require knowledge of a complete mathematical model that describes the relation between the system inputs and outputs, except perhaps for some parameters of the model that may need to be estimated from empirical data. Many of the physical models used are based on assumptions or approximations related to the noise (such as Markovianity) and the control (such as infinite bandwidth). We refer to the standard practice of fitting such models (i.e., finding the appropriate model parameters) given experimental data as a ``whitebox'' approach.

Finding an accurate model can be a challenging task especially for open quantum systems where the system has undesired interaction with its  environment and this environment is not a priori known. Therefore the task of identifying the system and its environment becomes a challenge on its own. Due to the difficulty of this task, unconventional methods have been developed such as machine learning (ML) methods. One ML approach is supervised learning \cite{ostaszewski2019approximation, khait2022optimal, zeng2020quantum}, where the common practice is to use abstract structures such as neural networks to model the map between a set of inputs and outputs. Another method is reinforcement learning, which is based on the idea of an agent navigating an unknown environment. The agent is rewarded if it approaches a target, and penalized if it hits an obstacle. This does not require prior information about the environment as the agent learns it during its operation. This is useful for quantum control applications \cite{niu2019universal,sivak2022model,baum2021experimental}, where the open system dynamics are treated as a blackbox, and the agent (controller) navigates through the state space to reach the control target. Both ML methods do not require knowing anything about the environment and do not give a set of physical dynamical equations, and thus they are considered ``blackbox'' methods. 

In situations where some physical insights are available, or the quantity of interest that is required for performing the control task is inaccessible experimentally, a hybrid approach is needed. We refer to this as a ``graybox'' approach. Part of the model uses blackbox structures such as neural networks, while other parts uses whitebox structures such as calculating a unitary given a Hamiltonian. This approach has been developed to model and control static (time-independent) unitary quantum photonic systems \cite{youssry2022experimental,youssry2020modeling} as well as open quantum systems subjected to classical environments \cite{youssry2020characterization,youssry2021noise}. It was also utilized for the characterization of superconducting qubits \cite{10.1103/prxquantum.2.040355} as well as geometric quantum gate synthesis \cite{perrier2020quantum}. In this paper, we consider the optimization of open-loop controls for single qubit gate synthesis when the a qubit  that is in contact with  an a priori unknown non-Markovian quantum environment (which is not a quantum white noise process), and analyze its performance. In order to obtain the controls, a graybox model based on an ML model proposed in \cite{youssry2020characterization} is used to approximate the unknown quantum noise. This graybox model is empirically identified based on measurement data collected from the qubit. It is then used as proxy for the unknown quantum noise and control pulses for the gate synthesis are optimized based on this model.

Quantum non-Markovian dynamics can be obtained by embedding them into a larger Markovian dynamics \cite{Luchnikov_2020}. This work considers non-Markovian environment models that can be embedded in a larger Markovian model containing an auxiliary system and a quantum white noise (or Markovian) bath. Such an embedding approach is well-known and has a long history in studies of classical (non-quantum) non-Markovian processes, see, e.g, \cite{Cox55}. In the quantum setting, a bosonic non-Markovian Gaussian bath with a non-flat spectral density can be approximated arbitrarily closely by coupling the system to a number of fictitious single mode quantum harmonic oscillators that are each in turn coupled to a bosonic Markovian bath and possess a Lorentzian spectral density \cite{Imamoglu94}. In a generalization of this basic idea, the fictitious quantum harmonic oscillators are often referred to as {\em pseudo-modes} \cite{DBG01,Mazzola09}. It was recently put forward that general Gaussian bosonic baths with an arbitrary spectral density can  be approximated by linear coupling to a (possibly infinite) number of quantum harmonic oscillators \cite{Mascherpa20}. A similar result has also been established for  non-Markovian fermionic baths with arbitrary spectral densities in \cite{CAG19} in which the bath can be approximated by coupling to a number of fermionic modes $a_k$ satisfying the anti-commutation relations $\{a_k,a_k^{\dag}\}=1$ to Markovian fermionic baths. 

After tracing out the white noise  bath, the joint reduced density  of the qubit and  auxiliary system undergoes an evolution according to a Lindblad master equation that  can be simulated using standard software packages. Tracing out the auxiliary Hilbert space from the joint density operator gives an evolution of the system density operator alone that no longer obeys a Lindblad master equation, as would be expected of a system coupled to a non-Markovian quantum noise. We use this method to simulate a  qubit coupled to a non-Markovian noise source and generate a dataset with which to train the ML models. We then show the results of using these empirically identified models to obtain the optimal control pulses for a set of targets using local and global optimization methods. We investigate both single axis and multi-axis control of the qubit.  

This paper is structured as follows. In section  \ref{sec:methods} we describe the system of interest and the setting of the control problem that is considered, give details of the non-Markovian noise model adopted and of the graybox ML model assumed for the unknown system, and how the ML model is trained and tested based on data. In section \ref{sec:results} we present the results of training the adopted non-Markovian model to the system of interest and the application of open-loop control designed based on the trained model based on the actual system. Finally, in section \ref{sec:conclu} we give a conclusion for this paper.

\section{Methods}
\label{sec:methods}
\subsection{System Dynamics}
The expectation of any observable $O$ for a closed quantum system initialized in the state $\rho(0)$ evolves over the time $[0,T]$ according to 
\begin{align}
    \mathbb{E}\{O(T)\}_\rho = \mathrm{tr}{ \left(U_{\rm ctrl}(T) \rho(0) U_{\rm ctrl}(T)^{\dagger} O\right)},
\end{align}
where the control unitary $U_{\rm ctrl}$ is given by the time-ordered evolution of the (generally time-dependent) system Hamiltonian $H_{\rm ctrl}$,
\begin{align}
    U_{\text{ctrl}}(T) = \mathcal{T}_+ e^{-i\int_0^T{H_{\text{ctrl}}(s) ds}},
    \label{equ:Uctrl}
\end{align}
where $\mathcal{T}_+$ denotes the time-ordering operator. For example, a qubit with a drift evolution along the $Z$-axis, and a control along the $X$- and $Y$- direction will have the Hamiltonian in the form
\begin{align}
    H_{\text{ctrl}}(t) = \frac{1}{2}\Omega \sigma_z + \frac{1}{2} f_x(t)  \sigma_x + \frac{1}{2} f_y(t) \sigma_y,
    \label{equ:Hctrl}
\end{align}
where $\Omega$ is the qubit energy gap, $\sigma_j$ is the Pauli operators along the $j^{\text{th}}$ direction, and $f_j(t)$ is the corresponding control pulse. We are interested in a qubit that is coupled to some non-Markovian noise that is in principle unknown. To explore the efficacy of the proposed ML based control method, we consider an environment quantum noise model that can be simulated by solving a Lindblad master equation, the details of the  modeling are given by in the next subsection. The non-Markovian noise model consists of an auxiliary quantum system that is coupled to a quantum white noise bath. The qubit, auxiliary and quantum white noise bath undergo a joint unitary evolution. 
By tracing out the white noise bath, the joint density operator of the qubit and auxiliary then evolve according to a master equation. The density operator for the qubit alone is then obtained by tracing out the auxiliary.  

If we are only interested in the observables of the system, then we can utilize the approach in \cite{youssry2020characterization} and write down the expectation of any observables for the system as 
\begin{align}
    \mathbb{E}\{O(T)\}_\rho = \mathrm{tr}{ \left(V_O(T) U_{\rm ctrl}(T) \rho(0) U_{\rm ctrl}(T)^{\dagger} O\right)}.
    \label{equ:exp}
\end{align}
The operator $V_O(T)$ captures all the information about the environment and how it affects the evolution of the system, and is given by
\begin{align}
    V_O(T) = \braket{O^{-1} \tilde{U}_I^{\dagger}(T) O \tilde{U}_I(T)},
\end{align}
with the expectation $\braket{\cdot} = \tr_{\text{env}}\left(\cdot \rho_{\text{env}}\right)$, with $\rho_{\rm env}$ being the initial state of the bath. $\tilde{U}_I$ is a modified interaction picture unitary that is defined in terms of the total unitary of system and bath $U(t)$ as  
\begin{align}
    U(t) = \tilde{U}_I(t) U_{\rm ctrl}(t).
\end{align}
The advantage of this formulation, is that it encodes all the relevant effects of the quantum environment (which is hard to simulate) in just one system operator, $V_O(T)$. It is independent of the noise bath model whether it is a classical bath, or a fermionic or bosonic bath, or even a combination of those. Additionally, with this structure, we can construct ML models that could be trained on actual experimental data, and used to estimate those noise operators and consequently do quantum control.

\subsection{Non-Markovian quantum noise models}\label{sec:Aux}

\begin{figure}[!ht]
\centering
\includegraphics[scale=0.5]{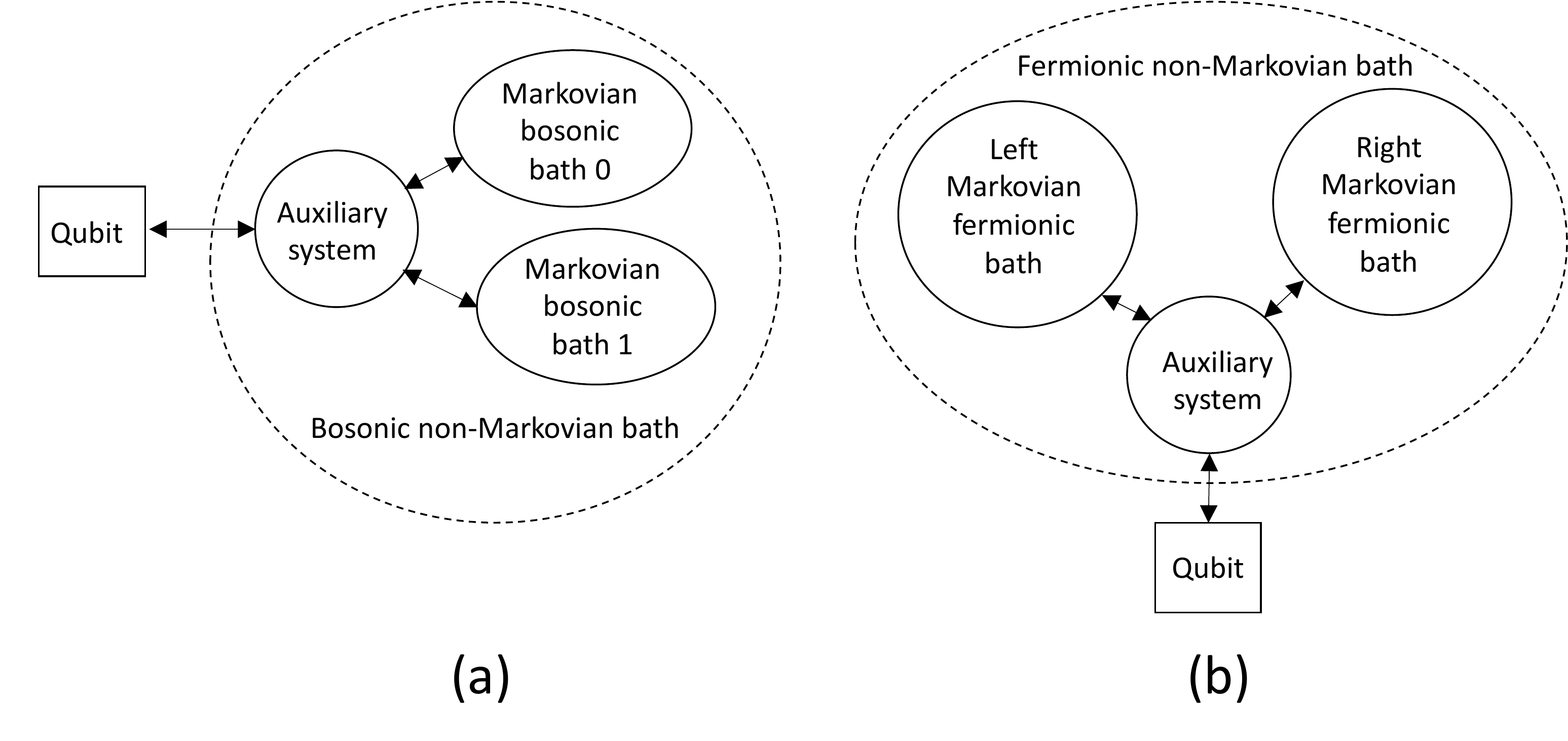}
\caption{A qubit coupled to (a) a bosonic non-Markovian bath and (b) a fermionic non-Markovian bath. Each non-Markovian bath is modeled as an auxiliary system that is coupled to two Markovian baths.}
\label{fig:non-Markovian_model}
\end{figure}

We now detail the bosonic and fermionic non-Markovian noise models that are described by single mode bosonic or fermionic operators which are coupled to a Markovian bosonic or fermionic bath, respectively; see Fig.~\ref{fig:non-Markovian_model}.  

\subsubsection{Qubit coupled to a Gaussian bosonic noise}
We begin by describing a qubit coupled to a non-Markovian bosonic Gaussian bath (we will use the word  quantum bath and quantum noise interchangeably). Let $H_{\rm ctrl}(t)$ be the (generally) time-dependent Hamiltonian of a qubit (as the system) and let $a$ be the annihilation operator of a quantum harmonic oscillator as an auxiliary system, satisfying the commutation relation $[a,a^{\dag}]=1$. The qubit is coupled to the quantum harmonic oscillator via the coupling $H_{\rm sa} = V\sigma_+ a +V^*\sigma_- a^{\dag}$, where $\sigma_+ = |1\rangle \langle 0|$ is the qubit raising operator and $\sigma_-=\sigma_+^{\dag}$ is the qubit lowering operator and $V$ is some complex coupling constant. The auxiliary is taken to have the Hamiltonian $H_{\rm a}=\omega_d a^{\dag}a$ and is in turn coupled to two Markovian baths in the vacuum state via the coupling (or jump) operators $L_0 = \sqrt{\gamma_0} a$ and $L_1 = \sqrt{\gamma_1} a^{\dag}$. The joint evolution of the combined  qubit-auxiliary-bosonic noise is given by a unitary propagator $U(t)$ given as the solution of the  Hudson-Parthasarathy quantum stochastic differential equation (QSDE) \cite{HP84,KRP92} (see also  \cite{CKS17} for a physics-oriented introduction):
\begin{align*}
dU(t)&=\left((-i(H_{\rm ctrl}(t) + H_{\rm a} + H_{\rm sa}) -(1/2)L_0^{\dag}L_0-(1/2)L_1^{\dag} L_1)dt+ L_0  dB^{\dag}_{0,t} + L_1 dB^{\dag}_{1,t} \right.\\
&\quad \left. - L_0^{\dag}dB_{0,t} - L_1^{\dag} dB_{1,t}\right)U(t),\;U(0)=I. 
\end{align*}
In the above $B_{j,t} = \int_{0}^t b_{j,s} ds$ and $B^{\dag}_{j,t} = \int_{0}^t b^{\dag}_{j,s} ds$ are the annihilation and creation operators of bosonic bath $j$ ($j=0,1$), and $b_{j,t}$ is a bosonic quantum white noise process satisfying the commutation relation $[b_{j,t},b^{\dag}_{k,\tau}]=\delta_{jk} \delta(t-\tau)$ for $j,k=0,1$. When the bosonic Markovian baths is in a vacuum (or coherent state), the infinitesimal increments $dB_{j,t} = b_{j,t} dt$ and  $dB^{\dag}_{j,t} = b_{j,t} dt$ satisfy the product rule \cite{HP84,KRP92} $dB_{j,t} dB_{k,t}^{\dag}=\delta_{jk}dt$ with all other products of $dB_{j,t}$ and $dB_{k,t}^{\dag}$ with itself or with the other being equal to 0 for $j,k=0,1$. The qubit and auxiliary is initialised in the product state $\rho_{\rm sa,0}=\rho_{\rm s,0} \otimes \rho_{\rm a,0}$, where $\rho_{\rm s,0}$ is the initial state of the qubit and $\rho_{\rm a,0}$ the initial state of the auxiliary. 

The reduced density operator of the qubit + auxiliary, after tracing out the bosonic bath, $\rho_{\rm sa}(t)$ is given by $\rho_{\rm sa}(t)=\mathrm{tr}_{\mathcal{H}_{B}}(U(t)^{\dag} \rho_{\rm sa,0} |0\rangle_B \langle 0| U(t))$, where $|0\rangle_B \langle 0|$ is the vacuum state of the Markovian bath and $\mathrm{tr}_{\mathcal{H}_{B}}$ denotes partial tracing over the Markovian bath, and satisfies the Lindblad master equation,
$$
\dot{\rho}_{\rm sa}(t) = i[\rho_{\rm sa}(t),H_{\rm ctrl}(t) + H_{\rm a} + H_{\rm sa}] + L \rho_{\rm sa}(t) L^{\dag} -\frac{1}{2}[L^{\dag}L, \rho_{sa}(t)]. 
$$
The reduced density operator equation for the qubit itself, $\rho_{\rm s}(t)$ is given by tracing out the auxiliary from $\rho_{\rm sa}(t)$, i.e., $\rho_{\rm s}(t)=\mathrm{tr}_{\mathcal{H}_A}(\rho_{\rm sa}(t))$ where $\mathrm{tr}_{\mathcal{H}_A}$ is a partial trace over the auxiliary, would not in general satisfy a Markovian master equation but some non-Markovian master equation. 

\subsubsection{Qubit coupled to a fermionic noise}

We now describe the non-Markovian noise model that describes a fermionic quantum environment. Here we have a  quantum dot as an auxiliary system that is coupled to two Markovian fermionic baths, a left ohmic bath and a right ohmic bath  \cite{Milburn00,GJGN11}. The left bath is at zero temperature with energy less than its chemical potential, so the channel is initially fully occupied with average electron number $N_e=1$. The right bath is also at zero temperature and has energy larger than its chemical potential, thus $N_e=0$. 

The quantum dot has fermionic annihilation and creation operators 
$c$ and $c^{\dag}$, respectively, satisfying the canonical anti-commutation relation $\{c,c^{\dag}\}=1$. The Hamiltonian of the quantum dot is $H_{\rm a}=\omega_d c^{\dag}c$ and it is coupled to the qubit via the coupling Hamiltonian
$$
H_{\rm sa} = V\sigma_+ c + V^* \sigma_-c^{\dag},
$$
where $V$ is a complex coupling constant. The joint evolution of the qubit, auxiliary (i.e., quantum dot) and the left and right baths are given by the fermionic Hudson-Parthasarathy fermionic QSDE \cite{AH84} given by:
\begin{align*}
dU(t) &= \left((-i(H_{\rm ctrl}(t) + H_{\rm a} + H_{\rm sa}) - (1/2) L_0^{\dag}L_0 - (1/2) L_1^{\dag}L_1) dt \right.\\
&\qquad \left. + dB_{0,t}^{\dag}L_0 -L_0^{\dag} dB_{0,t} + dB_{1,t}^{\dag}L_1  -L_1^{\dag} dB_{1,t}\right)U(t),\, U(0)=I \end{align*}
where $L_0 =\sqrt{\gamma_R} c$ and $L_1 = \sqrt{\gamma_L} c^{\dag}$ are the coupling operators of the quantum dot to the left and right Markovian fermionic baths for some positive real coupling rates $\gamma_R$ and $\gamma_L$, respectively. Here  $B_{j,t} =\int_{0}^t b_{j,s}ds$ and $B^{\dag}_{j,t} =\int_{0}^t b^{\dag}_{j,s}ds$ are the annihilation and creation operators for the left bath ($j=1)$ and right bath ($j=0$), and $b_{j,t}$ is fermionic white noise process satisfying the canonical anti-commutation relations $\{b_{j,t},b_{k,s}^{\dag}\}=\delta_{j,k} \delta(t-s)$. 

The reduced density operator of the system and auxiliary, $\rho_{sa}(t)$, as in the bosonic case is given by $\rho_{\rm sa}(t)=\mathrm{tr}_{\mathcal{H}_{B}}(U(t)^{\dag} \rho_{\rm sa,0} |0\rangle_B \langle 0| U(t))$, where $|0\rangle_B \langle 0|$ is the joint vacuum (zero temperature) state of the Markovian left and right fermionic baths, and $\mathrm{tr}_{\mathcal{H}_{B}}$ denotes partial tracing over the Markovian fermionic baths. It satisfies the quantum master equation,
$$
\dot{\rho}_{\rm sa}(t) = i[\rho_{\rm sa}(t),H_{\rm ctrl}(t) + H_{\rm a } + H_{\rm sa}] + L_0 \rho_{\rm sa}(t) L_0^{\dag} -\frac{1}{2}[L_0^{\dag}L_0, \rho_{\rm sa}(t)] + L_1 \rho_{\rm sa}(t) L_1^{\dag}  -\frac{1}{2}[L_1^{\dag}L_1, \rho_{\rm sa}(t)]. 
$$

\begin{remark}
More generally, one could have a non-Markovian bath that is composed of both bosonic and fermionic non-Markovian baths, with a simultaneous coupling to both bosonic and fermionic auxiliary systems \cite{HP86,GJGN11}. 
\end{remark}

\subsection{Machine Learning Model}
\begin{figure}
    \centering
    \includegraphics[scale=0.75]{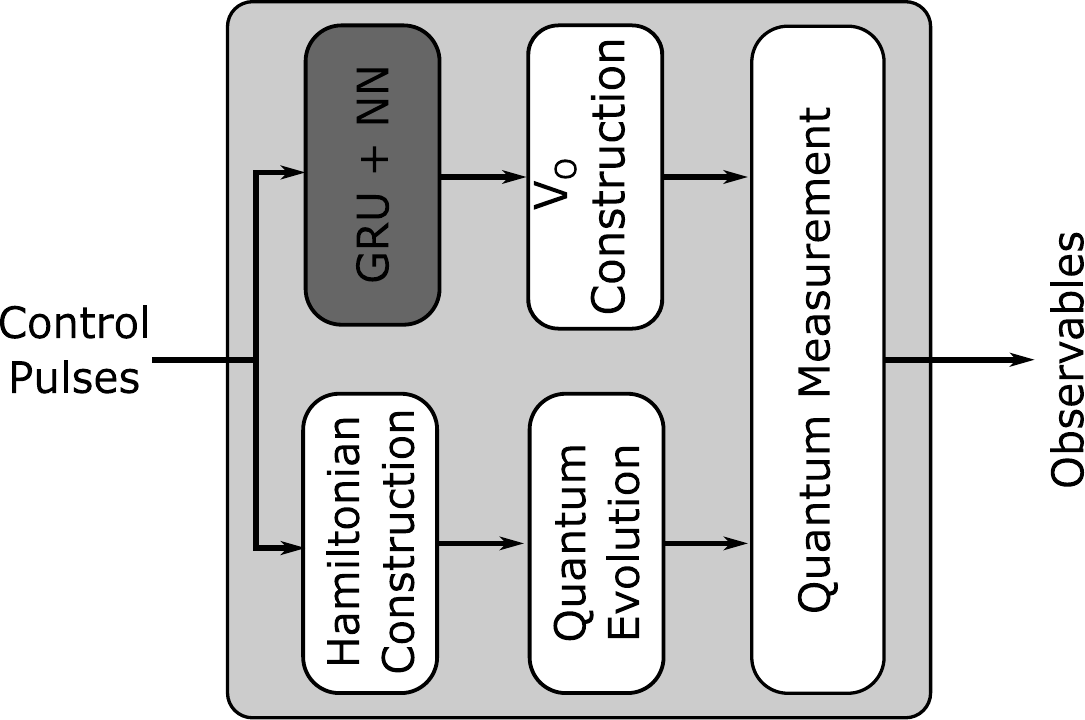}
    \caption{A schematic of the graybox model used in this paper. The inputs are the control pulses, and the outputs are the set of Pauli observables. Internally, a blackbox formed of Gated-Recurrent Units and Neural Networks estimates the parameters of the noise operators $V_O(T)$. The remaining layers are whitebox layers that computes the other physical quantities. We have computation of the Hamiltonian and unitary given the control pulses, the noise operators given their parameterization, and finally the expectations of observables given all other quantities.}
    \label{fig:model}
\end{figure}

We are interested in constructing an ML model that maps the control pulses acting on a quantum system to a set of expectation values of some system observables. Eq. \eqref{equ:exp} provides the necessary mathematical relation. We assume we do not know anything about the environment. In this case, as described earlier, the use of a whitebox approach for characterizing a quantum system has its limitations. It is difficult to express the noise operator $V_O(T)$ analytically without an approximation or a strong assumption that may not reflect the physical reality. On the other hand, using a fully blackbox approach to fit the measurement outcomes directly as a function of the controls could potentially provide accurate models, however, it might not be suitable for quantum control. The reason behind this is that optimal control will require accessing the noise operators $V_O(T)$. These are not accessible quantities in an experiment, so we cannot provide them as part of a dataset that an ML algorithm can directly learn. We can only provide a dataset of expectations of observables. A fully blackbox model would only predict these observables by learning an abstract representation that is only suitable for a ``machine''. Therefore, we utilize the graybox approach to take the advantage of both methods \cite{youssry2020modeling,youssry2020characterization,youssry2021noise, youssry2022experimental}. Fig. \ref{fig:model} shows the general structure of this model that implements eq. \eqref{equ:exp}. The inputs are the control pulses that we apply to the system, and the output would be the expectation of the system observables. Particularly, we choose an informationally-complete set of observables, such as the Pauli operators measured for all sets of Pauli eigenstates as initial states of the system. There are six possible eigenstates and three observables so this is a total of 18 outputs of the model. The ``black'' part of the graybox model is used to learn the parameters of the $V_O$ operators for each of the three observables as a function of the control pulses. These do not depend on the initial of the system. For this, we use two layers of Gated Recurrent Units (GRU) with 100 hidden nodes in each of them. The first layer takes the control pulses as an input. The second layer is then followed by a neural layer that generates the parameters of the matrix representation of each of the $V_O(T)$ operators. We do not care about the exact representation of the operators, we only need to be able to compute them, hence the use of abstract ML layers. Finally, a custom layer reconstructs the $V_O(T)$ operators given the parameters. The remaining parts of eq. \eqref{equ:exp} are known, and thus they are hard coded as the ``white'' part of the model. This includes the calculation of the closed-evolution unitary $U_{\rm ctrl}(T)$ as a function of the control pulses, the system evolution, and finally the measurement. This graybox structure solves the issue of a fully blackbox model. The outputs of the model are the expectations of observables which are accessible in experiment, and at the same time we have internal layers that could estimate the $V_O(T)$ operators. All the trainable parameters exist in those layers only. Therefore, for the model to be consistent with a dataset during training, the ``black'' part of the model is enforced to represent the $V_O(T)$ operators using the same representation of eq. \eqref{equ:exp}.
\subsection{Training, Testing, and Control}
In order to train and test the ML model, we need to construct datasets. Experimentally, we start by initializing the system in one of the Pauli eigenstates, apply a random pulse sequence, and then measure the three Pauli observables after system evolution. This is repeated for all the initial states of the system. The pair of a control pulse sequence and the set of observables constitute one example of our training dataset. We can then repeat the whole procedure, each time randomizing the control sequence, to construct the full dataset for training. The choice of the pulse shapes is usually dependent on the experimental setup as well as the quantum system itself. Practically, it will be constrained by some engineering limitations such as maximum power or bandwidth. Thus, it is convenient to parameterize the waveform in a way that makes it easy to express such constraints. For example, the waveform of a train of $N_p$ Gaussian pulses of finite width can be expressed as 
\begin{align}
    f(t) = \sum_{i=1}^{N_p} {A_i e^{-(t-\mu_i)^2/\sigma^2}},
    \label{equ:pulses}
\end{align}
where $A_i$ and $\mu_i$ are the amplitude and location of each pulse. If the waveform has finite power, then $A_i \in [A_{\text{min}}, A_{\text{max}}]$. This is will be important when it comes to performing quantum control.

Once we have this dataset, we can use the it to train the graybox model, where the trainable parameters of the model are optimized to minimize a loss function that measures the error between the model prediction and the actual output from the dataset. In this paper, we use the standard mean square error (MSE) averaged over the training set as a loss function, and we use the gradient-descent based ADAM optimizer \cite{Adam} for the training. At each iteration, we monitor the value of the loss function to see how well the model is performing in terms of prediction accuracy. The MSE curve versus the iteration number is referred to as the learning curve.

One important desirable behaviour of any ML algorithm is to ensure that it does not overfit, i.e., memorize the training set and not be able to predict outputs for new inputs. To assess this performance, we construct a ``testing'' dataset similar to the training dataset, and evaluate the MSE averaged over the testing set at each iteration. These testing examples do not affect the training procedure, they are only used for monitoring. A good model that does not overfit would have both its learning and testing curves close to each other, both converging to arbitrary small values. A related aspect is the generalization performance. In general, an ML algorithm will only generalize (i.e. predict outputs accurately) if tested against examples that are ``close'' to the training set. In our case, the model is expected to predict output of controls that have the same pulse shape as the training set. For example, a model trained with square pulses is expected to have good predictions for square pulses but not for Gaussian shaped pulses. While this could be a general limitation of the ML paradigm, in practical scenarios it is might not be of concern. Usually, the pulse shape is fixed from the beginning in an experimental setup. Thus, we do not need to obtain a model that generalizes to every possible input waveform. In a situation where we are interested to work with multiple waveform shapes, the training set has to include a balanced number of examples representing each of the pulse shapes of interest.

Once the model is trained, we can then use it to completely replace the system and act as a simulator for its dynamics. We can then use it as a part of an optimization loop to find optimal control in order to achieve a target. In this paper, we focus on implementing quantum gates, so there are two objectives: First to cancel the noise, which is equivalent to having $V_O(T) \to I$, where $I$ is the identity operator acting on the system. The second objective is to have the closed evolution unitary $U_{\rm ctrl}(T) \to G$, where $G$ is the target quantum gate. When we achieve this target, the expectation value of a system observable $O$ is given by $ \mathbb{E}\{O(T)\}_\rho = \mathrm{tr}{ \left(G \rho(0) G^{\dagger} O\right)}.$ On the other hand, from the trained model, we can estimate the expectation of observables given the control pulses as
\begin{align}
    \hat{\mathbb{E}}\{O(T)\}_\rho = \mathrm{tr}{ \left(\hat{V}_O(T) U_{\rm ctrl}(T) \rho(0) U_{\rm ctrl}(T)^{\dagger} O\right)},
\end{align}
where $\hat{V}_O(T)$ is the ML prediction of the noise operator given the control pulses. A suitable cost function for determining the control pulses to implement a desired gate $G$, which is aligned with how the model is trained, is taken to be of the form
\begin{align}
    J = \sum_{\rho(0), O}{\left(\mathrm{tr}{ \left(G \rho(0) G^{\dagger} O\right)} - \mathrm{tr}{ \left(\hat{V}_O(T) U_{\rm ctrl}(T) \rho(0) U_{\rm ctrl}(T)^{\dagger} O\right)}  \right)^2 },
\end{align}

and we try to find an optimal pulse sequence that  minimizes $J$. Notice that control affects both the unitary and the noise operators. 
The next step would then optimize $J$ to get the optimal control. In this paper, we explore two types of optimizers: gradient-descent and genetic algorithm. The genetic algorithm has an advantage of being more immune to falling in local minima at the expense of being computationally more expensive. The optimizer also has to take into consideration the constraints of the control. 

\section{Results}
\label{sec:results}
\subsection{Implementation}
In this paper, we performed a numerical study to assess the performance of the graybox ML model and its use for control on a single qubit system in the presence of a quantum bath. As discussed in Section \ref{sec:Aux}, we use an auxiliary quantum system to enable the simulation of the system dynamics under two cases. The first is when we have a fermionic environment, with the total system+auxiliary Hamiltonian in the form
\begin{align}
    H_{\text{sys + aux}}(t) = \frac{1}{2} \omega_s \sigma_z + \frac{1}{2} f_x(t) \sigma_x + \frac{1}{2} f_y(t) \sigma_y +  \omega_d c^{\dagger}c + V \sigma_{+} c + V^* \sigma_{-} c^{\dagger},  
\end{align}
where $c$ is the fermionic annihilation operator for the quantum dot. The Lindblad operators of the auxiliary are $\{ \sqrt{\gamma_L}c, \sqrt{\gamma_R}c^{\dagger} \}$. The fermionic creation and annihilation operators can be simulated by considering the Jordan-Wigner transformation which aims to express those operators in terms of standard qubit Pauli operators \cite{Spee18}. This means that we can simulate the combined system as a the evolution of finite number of qubits and thus is exact. Since we are considering a single-mode only, one qubit is sufficient to model the auxiliary. In this paper, we set the Hamiltonian parameters to be $\omega_s=12$, $\omega_d=5$, $V=2$, $\gamma_L=\gamma_R=0.7$. The initial state of the auxiliary is a fermionic thermal state with inverse temperature parameter $\beta=1$, and chemical potential $\mu=3$. 

The other case is for a bosonic bath. In this case the qubit is coupled to a quantum bosonic cavity with total system+auxiliary Hamiltonian in the form 
\begin{align}
    H_{\text{sys + aux}}(t) = \frac{1}{2} \omega_s \sigma_z + \frac{1}{2} f_x(t) \sigma_x + \frac{1}{2} f_y(t) \sigma_y + \omega_d a^{\dagger}a + V \sigma_{+} a + V^* \sigma_{-} a^{\dagger},  
\end{align}
where $a$ is the annihilation operator of the cavity. The Lindblad operators of the auxiliary are $\{ \sqrt{\gamma_L}a, \sqrt{\gamma_R}a^{\dagger} \}$. These operators are infinite-dimensional, and therefore the Hilbert space of the cavity has to be truncated to a some finite dimension to allow simulations. The dimension is chosen by trial and error where we keep increasing it and monitor the change in the values of the quantum measurements, until we reach a point where there is no more change. In this case, we set $\omega_s=12$, $\omega_d=5$, $V=1.3$, $\gamma_L=\gamma_R=0.7$. The dimension of the Hilbert space of the auxiliary is truncated to 20. The initial state of the system is a bosonic thermal state with average number of photons $n=1$.

We consider these single-mode auxiliaries arbitrarily. If we want to simulate a bath with some properties (such as a particular autocorrelation), then we need to find an optimal auxiliary system that would give rise to the same statistics. This is out of the scope of this paper, but we refer as an example to \cite{Mascherpa20,CAG19} for more details on how to perform this method. Most importantly, the purpose of these simulations is to create a dataset that simulates a qubit in a quantum bath. It is not part of the ML algorithm. The use of the graybox architecture together with the framework of eq. \eqref{equ:exp} is what enables us to express all the bath interaction as a single system operator regardless of the environment characteristics. Therefore, in an experimental setting, we do not need to simulate the evolution of the system, or consider the auxiliary system or the bath. We just control and measure the system directly in the experiment.

The time evolution is fixed to $T=1$, and is discretized into $M=1024$ time steps. The control takes the form of $N_p=5$ non-overlapping Gaussian pulses with $A_{\text{max}} = 25$ as in eq. \eqref{equ:pulses}. We fix the pulse width to the value $\sigma = T/(12N_p)$, while randomizing the amplitude and location of each pulse per each example. We also ensure that these pulses do not overlap with respect to each other for any given example in the dataset. For the single-axis datasets, we have pulses in along x-direction only (i.e. $f_y(t)=0$), while for the multi-axis datasets, we choose the waveform for each of the x- and y- directions. For the measurement statistics, we consider finite sampling, so we study three cases: $N=512$, $N=1024$, and $N=\infty$ which represents the ideal case. We then construct 9000 examples for training and 1000 for testing for each case (bath type, number of controls, and number of shots). This gives a total of 12 generated datasets for this set of experiments.

\begin{figure}[]
    \centering
    \includegraphics[scale=0.5]{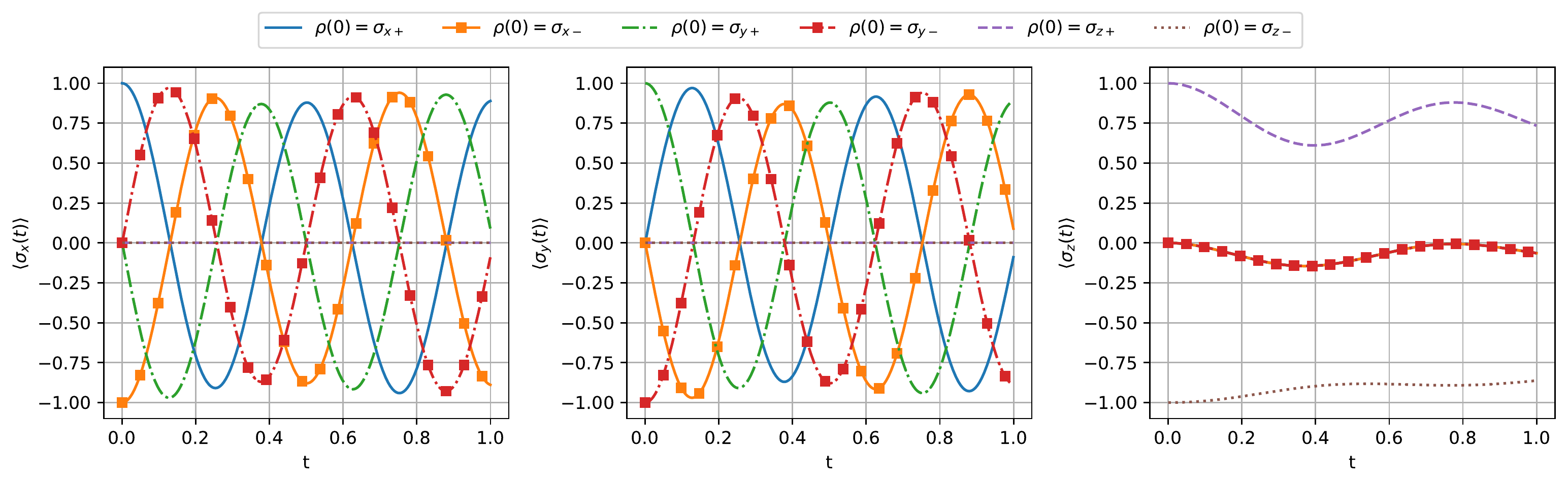}
    \caption{The expectations of the Pauli observables of the qubit under free evolution in the fermionic bath as a function of time. The initial state is one of the Pauli eigenstates. The $\sigma_{x}$ and $\sigma_y$ expectations of the $\sigma_{z\pm}$ states vanish, while the $\sigma_z$ expectation value of the $\sigma_{x\pm}$ and $\sigma_{y\pm}$ states coincide.}
    \label{fig:F_expectations}
\end{figure}

\begin{figure}[]
    \centering
    \includegraphics[scale=0.5]{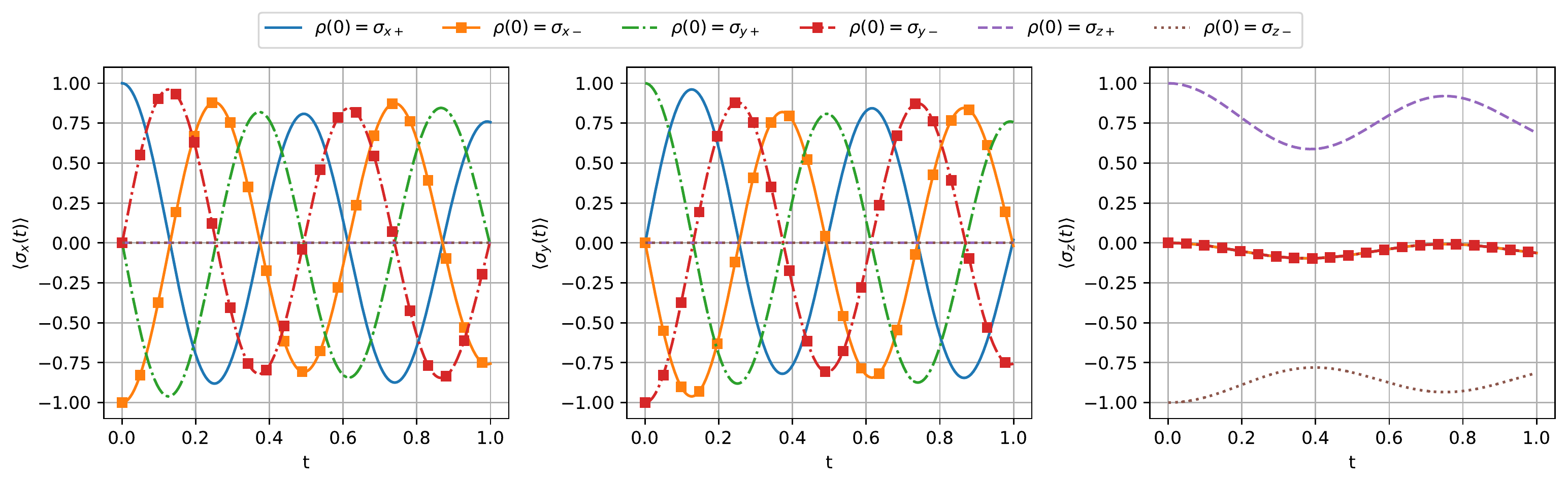}
    \caption{The expectations of the Pauli observables of the qubit under free evolution in the bosonic bath as a function of time.The initial state is one of the Pauli eigenstates. The $\sigma_{x}$ and $\sigma_y$ expectations of the $\sigma_{z\pm}$ states vanish, while the $\sigma_z$ expectation value of the $\sigma_{x\pm}$ and $\sigma_{y\pm}$ states coincide.}
    \label{fig:B_expectations}
\end{figure}

\begin{figure}
    \centering
    \includegraphics[scale=0.5]{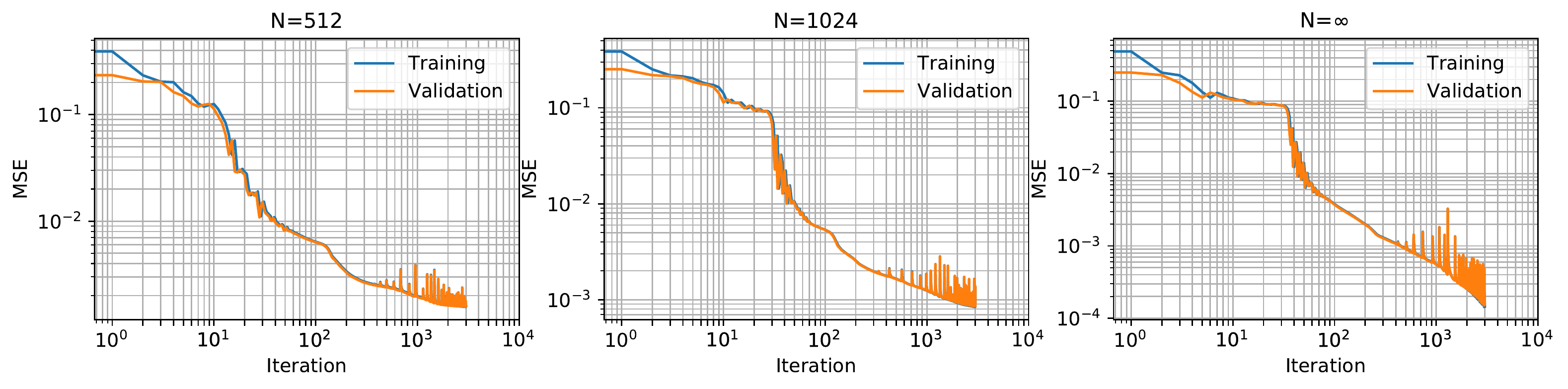}
    \caption{Evaluation of graybox model training and testing for single-axis control applied to the system in the fermionic bath }
    \label{fig:F_SA}
\end{figure}

\begin{figure}
    \centering
    \includegraphics[scale=0.5]{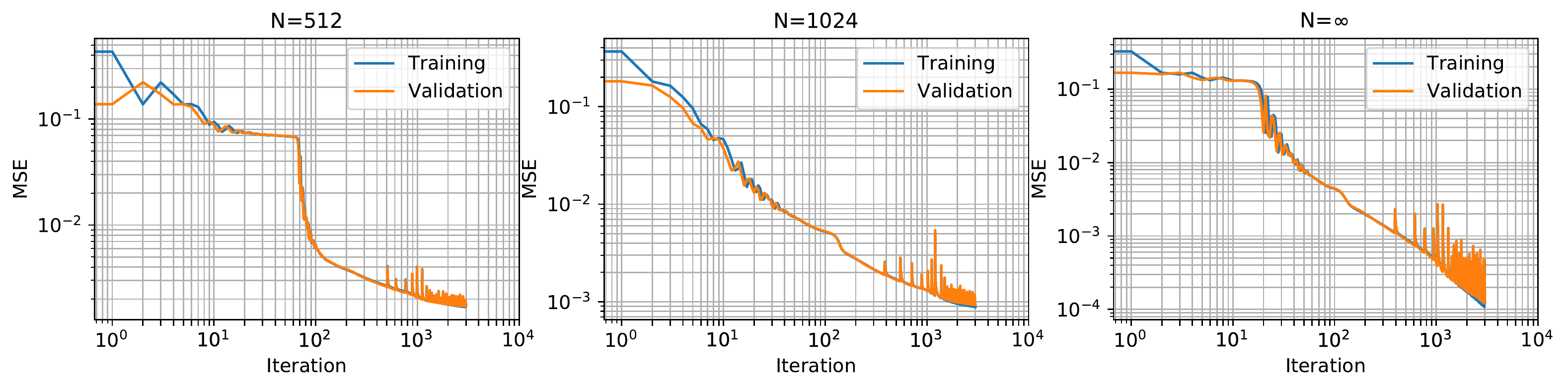}
    \caption{Evaluation of graybox model training and testing for single-axis control applied to the system in the bosonic bath }
    \label{fig:B_SA}
\end{figure}

\begin{figure}
    \centering
    \includegraphics[scale=0.5]{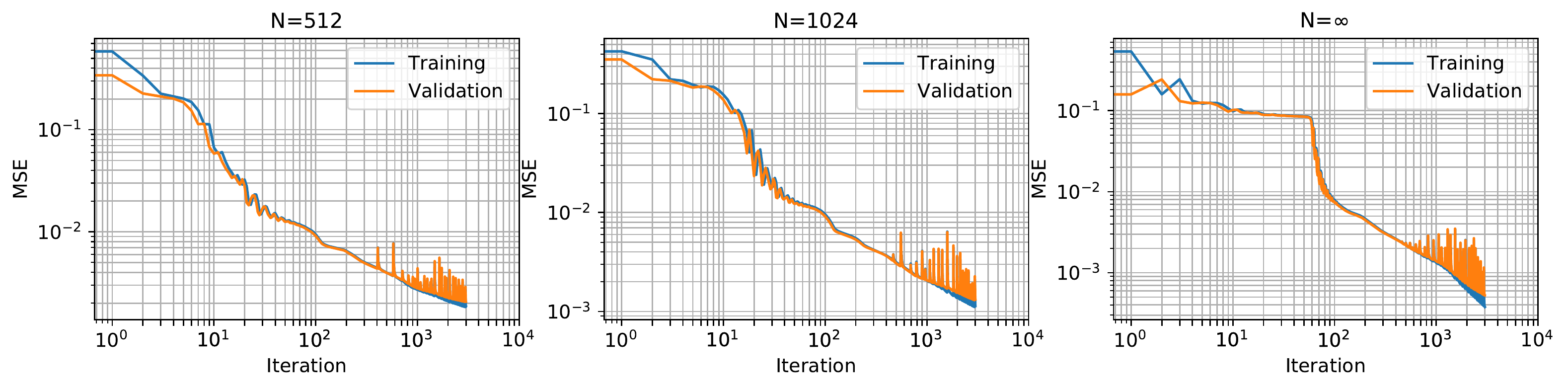}
    \caption{Evaluation of graybox model training and testing for multi-axis control applied to the system in the fermionic bath }
    \label{fig:F_MA}
\end{figure}

\begin{figure}
    \centering
    \includegraphics[scale=0.5]{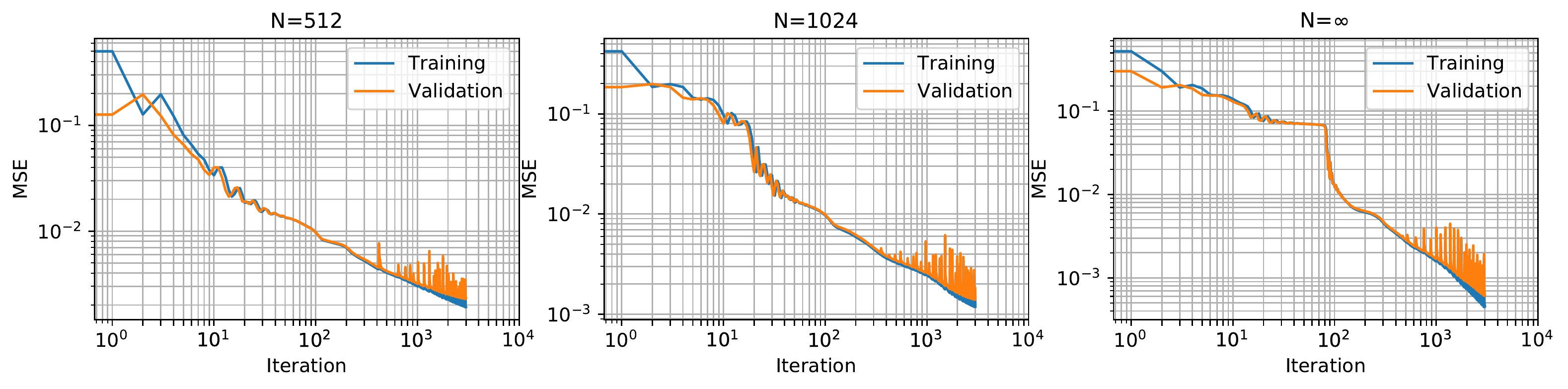}
    \caption{Evaluation of graybox model training and testing for multi-axis control applied to the system in the bosonic bath }
    \label{fig:B_MA}
\end{figure}

We used the QuTip Python package \cite{johansson2012qutip, JOHANSSON20131234} to solve the Lindblad master equation for the combined system and auxiliary Hamiltonians, and then the system observables are evaluated. In Figs. \ref{fig:F_expectations},\ref{fig:B_expectations}, we show the results of simulating the dynamics of the qubit under free evolution (i.e., in the absence of control) for the fermionic and bosonic baths, respectively. We show the expectations of the three Pauli observables, taking every Pauli eigenstate as initial state. We then construct the training and testing datasets as described earlier. After constructing the datasets, we implement the ML model using the Tensorflow Python package \cite{tensorflow, keras}. We train the graybox model and assess its performance on each of the 12 datasets separately (i.e., we obtain 12 separate models). Figs. \ref{fig:F_SA}, \ref{fig:B_SA} show the results of the training for single-axis control for fermionic and bosonic baths, while Figs. \eqref{fig:F_MA}, \eqref{fig:B_MA} show that for the multi-axis control. Table \ref{tab:mse} summarizes the final MSE obtained at the end of the training iterations. 

Next, we utilize the trained models to design the optimal control pulses in order implement a universal set of quantum gates: The identity gate $I$, the three Pauli gates $X,Y,Z$, the Hadamard gate $H$, and the rotation about X-axis gate $R_X(\pi/4)$. The optimal control pulses are then tested by simulating the evolution of the pulse-driven system in the quantum bath. We report the process fidelity in Tables \ref{tab:F_SA_P}, \ref{tab:B_SA_P}, \ref{tab:F_MA_P}, \ref{tab:B_MA_P}, as a metric to assess the performance of our methods. The evaluation of these metrics requires the calculation of the Choi state at the end of the evolution time interval \cite{graphical}. This is done by simulating the evolution of two copies of the system starting from an EPR initial state, where one half of the pair undergoes the evolution alongside the auxiliary, while the other half remains unchanged. The evolved state of the system after tracing out the auxiliary is the Choi state up to a normalization factor. It is important to note that this simulation requires knowledge of the bath, but it is only used to assess the algorithm performance. It is not required for the control procedures. In an experimental setting, we can do process tomography for these six gates to asses the performance. We perform the control assessment for both the gradient-descent as well as genetic algorithm optimizers.
 
\begin{table}[]
\begin{tabular}{|l|cc|cc|}
\hline
\multirow{2}{*}{}                 & \multicolumn{2}{c|}{\textbf{Fermionic}}                        & \multicolumn{2}{c|}{\textbf{Bosonic}}                          \\ \cline{2-5} 
                                  & \multicolumn{1}{c|}{Training} & Testing               & \multicolumn{1}{c|}{Training} & Testing               \\ \hline
\multicolumn{1}{|c|}{\textbf{Single-axis}} & \multicolumn{1}{c|}{} & & \multicolumn{1}{c|}{}&   \\ 
\multicolumn{1}{|c|}{\quad $N=512$} & \multicolumn{1}{c|}{$1.58\times 10^{-3}$} & $1.60\times 10^{-3}$ & \multicolumn{1}{c|}{$1.72\times 10^{-3}$} & $1.76\times 10^{-3}$  \\ 
\multicolumn{1}{|c|}{\quad $N=1024$} & \multicolumn{1}{c|}{$9.44\times 10^{-4}$} & $8.70\times 10^{-4}$ & \multicolumn{1}{c|}{$8.83\times 10^{-4}$} & $9.28\times 10^{-4}$  \\ 
\multicolumn{1}{|c|}{\quad $N=\infty$} & \multicolumn{1}{c|}{$1.45\times 10^{-4}$} & $1.55\times 10^{-4}$ & \multicolumn{1}{c|}{$1.24\times 10^{-4}$} & $1.54\times 10^{-4}$  \\ 
\multicolumn{1}{|c|}{\textbf{Multi-axis}} & \multicolumn{1}{c|}{} & & \multicolumn{1}{c|}{}&   \\ 
\multicolumn{1}{|c|}{\quad $N=512$} & \multicolumn{1}{c|}{$1.85\times 10^{-3}$} & $2.07\times 10^{-3}$ & \multicolumn{1}{|c|}{$1.91\times 10^{-3}$} & $2.29\times 10^{-3}$  \\ 
\multicolumn{1}{|c|}{\quad $N=1024$} & \multicolumn{1}{c|}{$1.14\times 10^{-3}$} & $1.32\times 10^{-3}$ & \multicolumn{1}{|c|}{$1.19\times 10^{-3}$} & $1.44\times 10^{-3}$ \\ 
\multicolumn{1}{|c|}{\quad $N=\infty$} & \multicolumn{1}{c|}{$3.77\times 10^{-4}$} & $5.23\times 10^{-4}$ & \multicolumn{1}{|c|}{$4.48\times 10^{-4}$} & $6.13\times 10^{-4}$ \\ 
 \hline
\end{tabular}
\caption{The final MSE evaluated for the model training and the testing over the various datasets}
\label{tab:mse}
\end{table}

\begin{table}[h]
\begin{tabular}{|c|ccc|ccl|}
\hline
\multirow{2}{*}{Gate}           & \multicolumn{3}{c|}{Gradient-Descent}                                   & \multicolumn{3}{c|}{Genetic Algorithm}                                                       \\ \cline{2-7} 
                                & \multicolumn{1}{c|}{$N=512$} & \multicolumn{1}{c|}{$N=1024$} & $N=\infty$ & \multicolumn{1}{c|}{$N=512$} & \multicolumn{1}{c|}{$N=1024$} & \multicolumn{1}{c|}{$N=\infty$} \\ \hline
$I$&\multicolumn{1}{c|}{94.81}&\multicolumn{1}{c|}{94.69}&\multicolumn{1}{c|}{94.78}&\multicolumn{1}{c|}{94.39}&\multicolumn{1}{c|}{94.40}&\multicolumn{1}{c|}{94.43}\\ \hline
$X$&\multicolumn{1}{c|}{80.10}&\multicolumn{1}{c|}{84.82}&\multicolumn{1}{c|}{84.82}&\multicolumn{1}{c|}{81.22}&\multicolumn{1}{c|}{81.39}&\multicolumn{1}{c|}{81.86}\\ \hline
$Y$&\multicolumn{1}{c|}{90.68}&\multicolumn{1}{c|}{83.56}&\multicolumn{1}{c|}{90.67}&\multicolumn{1}{c|}{86.67}&\multicolumn{1}{c|}{87.51}&\multicolumn{1}{c|}{86.49}\\ \hline
$Z$&\multicolumn{1}{c|}{36.22}&\multicolumn{1}{c|}{36.64}&\multicolumn{1}{c|}{22.38}&\multicolumn{1}{c|}{41.60}&\multicolumn{1}{c|}{41.10}&\multicolumn{1}{c|}{41.00}\\ \hline
$H$&\multicolumn{1}{c|}{72.91}&\multicolumn{1}{c|}{76.94}&\multicolumn{1}{c|}{76.93}&\multicolumn{1}{c|}{73.50}&\multicolumn{1}{c|}{73.07}&\multicolumn{1}{c|}{74.34}\\ \hline
$R_X(\pi/4)$&\multicolumn{1}{c|}{95.09}&\multicolumn{1}{c|}{95.09}&\multicolumn{1}{c|}{95.07}&\multicolumn{1}{c|}{95.03}&\multicolumn{1}{c|}{95.04}&\multicolumn{1}{c|}{95.04}\\ \hline
\end{tabular}
\caption{The results of the process fidelity for single-axis control applied to the system in the fermionic bath}
\label{tab:F_SA_P}
\end{table}

\begin{table}[h]
\begin{tabular}{|c|ccc|ccl|}
\hline
\multirow{2}{*}{Gate}           & \multicolumn{3}{c|}{Gradient-Descent}                                   & \multicolumn{3}{c|}{Genetic Algorithm}                                                       \\ \cline{2-7} 
                                & \multicolumn{1}{c|}{$N=512$} & \multicolumn{1}{c|}{$N=1024$} & $N=\infty$ & \multicolumn{1}{c|}{$N=512$} & \multicolumn{1}{c|}{$N=1024$} & \multicolumn{1}{c|}{$N=\infty$} \\ \hline
                                
$I$&\multicolumn{1}{c|}{90.83}&\multicolumn{1}{c|}{90.89}&\multicolumn{1}{c|}{90.90}&\multicolumn{1}{c|}{90.78}&\multicolumn{1}{c|}{90.75}&\multicolumn{1}{c|}{90.80}\\ \hline
$X$&\multicolumn{1}{c|}{83.31}&\multicolumn{1}{c|}{83.31}&\multicolumn{1}{c|}{83.35}&\multicolumn{1}{c|}{80.27}&\multicolumn{1}{c|}{80.94}&\multicolumn{1}{c|}{80.26}\\ \hline
$Y$&\multicolumn{1}{c|}{88.56}&\multicolumn{1}{c|}{80.58}&\multicolumn{1}{c|}{88.54}&\multicolumn{1}{c|}{85.40}&\multicolumn{1}{c|}{85.56}&\multicolumn{1}{c|}{85.84}\\ \hline
$Z$&\multicolumn{1}{c|}{51.24}&\multicolumn{1}{c|}{53.11}&\multicolumn{1}{c|}{51.35}&\multicolumn{1}{c|}{46.58}&\multicolumn{1}{c|}{46.79}&\multicolumn{1}{c|}{46.21}\\ \hline
$H$&\multicolumn{1}{c|}{76.40}&\multicolumn{1}{c|}{76.83}&\multicolumn{1}{c|}{76.83}&\multicolumn{1}{c|}{72.77}&\multicolumn{1}{c|}{72.19}&\multicolumn{1}{c|}{71.75}\\ \hline
$R_X(\pi/4)$&\multicolumn{1}{c|}{91.29}&\multicolumn{1}{c|}{91.29}&\multicolumn{1}{c|}{91.31}&\multicolumn{1}{c|}{91.26}&\multicolumn{1}{c|}{91.28}&\multicolumn{1}{c|}{91.24}\\ \hline

\end{tabular}
\caption{The results of the process fidelity for single-axis control applied to the system in the bosonic bath}
\label{tab:B_SA_P}
\end{table}

\begin{table}[h]
\begin{tabular}{|c|ccc|ccl|}
\hline
\multirow{2}{*}{Gate}           & \multicolumn{3}{c|}{Gradient-Descent}                                   & \multicolumn{3}{c|}{Genetic Algorithm}                                                       \\ \cline{2-7} 
                                & \multicolumn{1}{c|}{$N=512$} & \multicolumn{1}{c|}{$N=1024$} & $N=\infty$ & \multicolumn{1}{c|}{$N=512$} & \multicolumn{1}{c|}{$N=1024$} & \multicolumn{1}{c|}{$N=\infty$} \\ \hline
$I$&\multicolumn{1}{c|}{95.07}&\multicolumn{1}{c|}{95.35}&\multicolumn{1}{c|}{95.69}&\multicolumn{1}{c|}{94.81}&\multicolumn{1}{c|}{94.63}&\multicolumn{1}{c|}{94.87}\\ \hline
$X$&\multicolumn{1}{c|}{93.62}&\multicolumn{1}{c|}{93.52}&\multicolumn{1}{c|}{94.22}&\multicolumn{1}{c|}{92.52}&\multicolumn{1}{c|}{92.68}&\multicolumn{1}{c|}{92.23}\\ \hline
$Y$&\multicolumn{1}{c|}{93.51}&\multicolumn{1}{c|}{93.82}&\multicolumn{1}{c|}{93.60}&\multicolumn{1}{c|}{91.38}&\multicolumn{1}{c|}{92.29}&\multicolumn{1}{c|}{92.22}\\ \hline
$Z$&\multicolumn{1}{c|}{82.20}&\multicolumn{1}{c|}{88.31}&\multicolumn{1}{c|}{88.64}&\multicolumn{1}{c|}{80.85}&\multicolumn{1}{c|}{80.04}&\multicolumn{1}{c|}{80.37}\\ \hline
$H$&\multicolumn{1}{c|}{93.86}&\multicolumn{1}{c|}{90.84}&\multicolumn{1}{c|}{92.01}&\multicolumn{1}{c|}{92.65}&\multicolumn{1}{c|}{92.38}&\multicolumn{1}{c|}{92.60}\\ \hline
$R_X(\pi/4)$&\multicolumn{1}{c|}{94.99}&\multicolumn{1}{c|}{95.52}&\multicolumn{1}{c|}{95.29}&\multicolumn{1}{c|}{95.07}&\multicolumn{1}{c|}{94.95}&\multicolumn{1}{c|}{95.20}\\ \hline
\end{tabular}
\caption{The results of the process fidelity for multi-axis control applied to the system in the fermionic bath}
\label{tab:F_MA_P}
\end{table}

\begin{table}[h]
\begin{tabular}{|c|ccc|ccl|}
\hline
\multirow{2}{*}{Gate}           & \multicolumn{3}{c|}{Gradient-Descent}                                   & \multicolumn{3}{c|}{Genetic Algorithm}                                                       \\ \cline{2-7} 
                                & \multicolumn{1}{c|}{$N=512$} & \multicolumn{1}{c|}{$N=1024$} & $N=\infty$ & \multicolumn{1}{c|}{$N=512$} & \multicolumn{1}{c|}{$N=1024$} & \multicolumn{1}{c|}{$N=\infty$} \\ \hline
$I$&\multicolumn{1}{c|}{90.53}&\multicolumn{1}{c|}{92.35}&\multicolumn{1}{c|}{91.98}&\multicolumn{1}{c|}{90.44}&\multicolumn{1}{c|}{90.64}&\multicolumn{1}{c|}{90.34}\\ \hline
$X$&\multicolumn{1}{c|}{90.41}&\multicolumn{1}{c|}{91.70}&\multicolumn{1}{c|}{91.00}&\multicolumn{1}{c|}{90.40}&\multicolumn{1}{c|}{90.08}&\multicolumn{1}{c|}{90.04}\\ \hline
$Y$&\multicolumn{1}{c|}{90.56}&\multicolumn{1}{c|}{91.49}&\multicolumn{1}{c|}{91.74}&\multicolumn{1}{c|}{89.47}&\multicolumn{1}{c|}{90.09}&\multicolumn{1}{c|}{90.08}\\ \hline
$Z$&\multicolumn{1}{c|}{84.54}&\multicolumn{1}{c|}{85.71}&\multicolumn{1}{c|}{81.09}&\multicolumn{1}{c|}{78.78}&\multicolumn{1}{c|}{79.30}&\multicolumn{1}{c|}{79.11}\\ \hline
$H$&\multicolumn{1}{c|}{90.29}&\multicolumn{1}{c|}{91.45}&\multicolumn{1}{c|}{88.47}&\multicolumn{1}{c|}{87.78}&\multicolumn{1}{c|}{87.55}&\multicolumn{1}{c|}{87.71}\\ \hline
$R_X(\pi/4)$&\multicolumn{1}{c|}{92.63}&\multicolumn{1}{c|}{93.17}&\multicolumn{1}{c|}{92.79}&\multicolumn{1}{c|}{91.60}&\multicolumn{1}{c|}{91.73}&\multicolumn{1}{c|}{91.57}\\ \hline
\end{tabular}
\caption{The results of the process fidelity for multi-axis control applied to the system in the bosonic bath}
\label{tab:B_MA_P}
\end{table}

Besides this set of experiments, we also explore the effect of the coupling strength between the qubit and the auxiliary system (i.e., the strength of coupling to the bath given by $V$), on the performance of the algorithm. For this, we generate 4 more datasets with $V=0.2$ and $V=1$ for the fermionic bath, and $V=0.13$ and $V=0.65$ for the bosonic bath. We compare against the previous case of $V=2$ and $V=1.3$ for fermionic and bosonic baths respectively. We use multi-axis control, fixing the number of shots to $N=1024$, and use the gradient-descent optimization for pulse design. In Figs. \ref{fig:F_purity} and \ref{fig:B_purity} we show the simulated dynamics of the qubit state purity under free evolution, varying the coupling strength for the two quantum baths. The initial state is taken to be the eigenstate of the Pauli operators. In the plot, the states $\sigma_{x+}, \sigma_{x-}, \sigma_{y+}$, and $\sigma_{y-}$ result in the same dynamics. However, the $\sigma_{z+}$ and $\sigma_{z-}$ give rise to different dynamics. We then run the same steps of the algorithm and we show in Figs. \ref{fig:F_MA_V} and \ref{fig:B_MA_V} the results of the model training and validation for the two baths. In table \ref{tab:compareV}, we compare the process fidelity for the same set of quantum gates.

\subsection{Discussion}

In this paper, we explore a quantum control method for open quantum systems in a scenario where a qubit is coupled to a quantum environment. We study different aspects of the proposed method such as single- versus multi-axis control, the effect of finite sampling and the choice of optimization algorithm, and the effect of the coupling strength. This is different from the work in \cite{youssry2020characterization} where the qubit is subjected only to classical stochastic noise. In that work, the focus was on the characterization of the system and how to construct the graybox model. Control was shown only for single-axis using the gradient-descent approach, and ideal quantum measurements were assumed. Moreover, the cost function for the control pulse design in \cite{youssry2020characterization} is different from the one in this work. The problem setting in \cite{youssry2022experimental} and \cite{youssry2020modeling} is also very different. There, a photonic system is considered, which is modelled through a static time-independent Hamiltonian with unknown dependence on control. Moreover, the assumption is that the system of interest is closed (i.e. only unitary evolution). By performing quantum measurements, the map between controls and Hamiltonian is learnt, through the graybox, and then utilized to design quantum gates.

\begin{figure}[]
    \centering
    \includegraphics[scale=0.5]{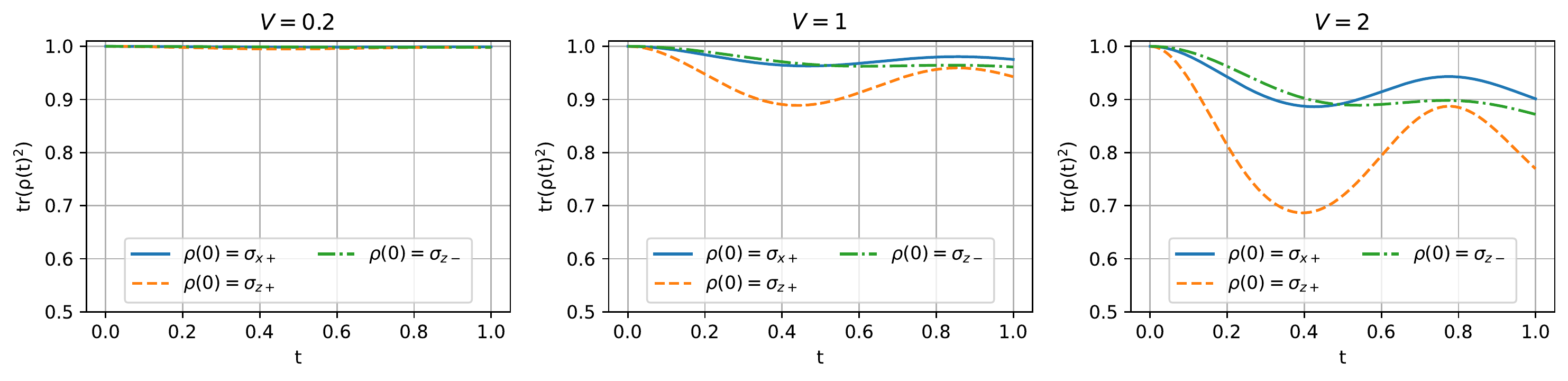}
    \caption{The quantum state purity of the qubit under free evolution in the fermionic bath as a function of time for different values of the coupling strength. The initial state is one of the Pauli eigenstates.}
    \label{fig:F_purity}
\end{figure}

\begin{figure}[]
    \centering
    \includegraphics[scale=0.5]{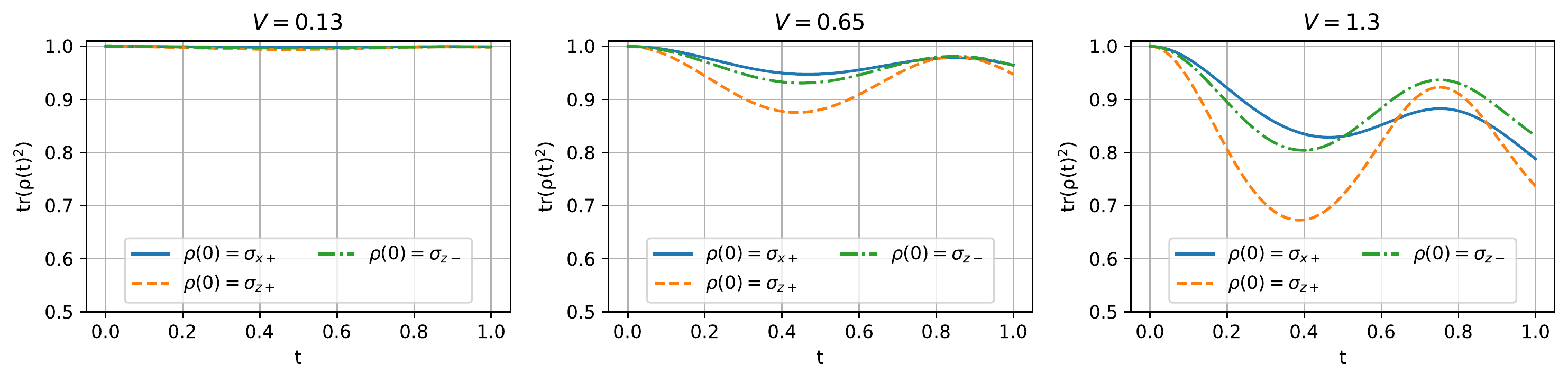}
    \caption{The quantum state purity of the qubit under free evolution in the bosonic bath as a function of time for different values of the coupling strength. The initial state is one of the Pauli eigenstates.}
    \label{fig:B_purity}
\end{figure}

The choice of our system and environment models as well as the simulation parameters in this paper showcases non-trivial system dynamics as in Figs. \ref{fig:F_expectations} and \ref{fig:B_expectations}. We see that all the Pauli eigenstates are subjected to decoherence but with different dynamics. For example, for the $\sigma_z$ observable, the $\sigma_{z+}$ and $\sigma_{z-}$ states have different trajectories, which are also different from the other eigenstates who have similar dynamics. The same classification of different dynamics exist as well for the evolution of the purity of the system state in Figs. \ref{fig:F_purity} and \ref{fig:B_purity}. So, a qubit initialized to a $\ket{0}$ would not remain in a pure state under this kind of quantum bath. This behaviour is different from the usual single-axis dephasing model which can take, for example, the form $H(t)=H_{\rm ctrl}(t) + \beta(t) \sigma_z$, where $\beta(t)$ is a classical stationary Gaussian random process, (see e.g, \cite{mohamed2020characterization} for further details). In that case, the Pauli states $\sigma_{x+}$, $\sigma_{x-}$,$\sigma_{y+}$, and $\sigma_{y-}$ lose purity over time, but the $\sigma_{z+}$ and $\sigma_{z-}$ would stay unaffected.

The plots in Figs. \ref{fig:F_SA} and \ref{fig:B_SA} show that the MSE of the training is decreasing with the increase of the number of iterations, indicating the ability of the model to learn the dataset. The MSE of the testing dataset also decreases with iterations. Additionally, it almost matches the training MSE indicating that the model does not overfit, which is an important desired feature of the model. In other words, the model does not just memorize the training set, but can also predict new examples without losing accuracy. This is also critical for the controller, since an open-loop controller will depend on the accuracy of the model. 

For the multi-axis datasets in Figs. \ref{fig:F_MA} and \ref{fig:B_MA}, we see a similar performance in terms of training and testing. However, comparing to the single-axis datasets, the single-axis model is performing better than the multi-axis. This is expected as there are more inputs and so the model has to learn more information. In this paper, we do not focus on optimizing the hyper-parameters of the model, but in  general the structures could be further optimized to enhance the performance. Also, increasing the dataset size would enhance the performance, but we limit it here to a size that can be feasible in terms of data collection time, for an experimental setup such as a superconducting qubit. 

Exploring the results of the controller, we find that multi-axis control outperforms single-axis control. We would expect this behaviour given the system dynamics we described earlier. The behaviour of the system deviates significantly from a unitary evolution and also from a simple dephasing noise. As a result, we do not expect that with just the drift term along the Z-axis and a constrained control along the X-axis, we can achieve any arbitrary gate (as the case with closed quantum systems, or open quantum system with dynamical decoupling). Having the extra control along the Y-axis improves the performance. Moreover, even though the single-axis trained models were better performing (lower MSE than multi-axis ones), the controller performance was actually better in the multi-axis case. This shows that the controller has some robustness against the errors in model, and that constraints imposed on the control pulses (such as limited power, bandwidth, and evolution time) affects the control performance more significantly. For most gates, we get high fidelities around 90\% when using the gradient-descent optimizer. The gradient-descent method and the genetic algorithm performed almost similarly, with gradient-descent slightly better in most cases. This indicates that even though the gradient-descent optimization suffers from the possibility of falling into a local minimum, the performance of the control was not affected when compared to global optimization with genetic algorithm. In general, the gradient-descent method is more efficient computationally, so it would be preferable to utilize if there is no significant degradation in performance compared to genetic optimization. Finally, we note that the performance of these optimization methods would depend on the various hyperparameters (such as the starting point, the learning rate, the number of iterations, genetic algorithm operations, etc.) So, an exact comparison is difficult. However, in an experimental setting with a specific target (rather than a general study), these hyperparameters would be optimized to acheive the best possible performance compared to the state-of-the-art method utilized for that platform.

Regarding the effect of statistical noise that occurs as a result of finite sampling, we see that increasing the number of shots will increase accuracy of the model (lower MSE). As shown in Table \ref{tab:mse}, this holds for the fermionic and bosonic datasets for both single-and multi-axis cases. This is an expected behaviour of the ML structure in response to the deviation of the quantum measurements from their ideal values. With regards to the effect of finite number of shots on control, Tables \ref{tab:F_SA_P},\ref{tab:B_SA_P},\ref{tab:F_MA_P}, \ref{tab:B_MA_P} show comparable performance of the gates utilizing the different models that were trained on the different datasets. Some gates performed better with noisier models, and other performed worse. This can be explained by looking back into the MSE performance of each model. We see that there is large order of difference in MSE when we increase the number of shots from finite to infinite, almost just one order of magnitude of enhancement. Thus, with the use of fidelity-based metrics, 
this enhancement in the MSE might not be sufficient to show significant change in the performance of different controllers. In other words, in terms of control, the models corresponding to different number of shots have almost the same accuracy, even though in terms of MSE they differ. This could be beneficial as constructing datasets in an experimental setting can be time consuming, especially for large sizes. If the most important desired aspect is finding optimal control, then a small number of shots might be sufficient. Finally, note that the performance of the gates was evaluated utilizing the exact simulator (with no sampling noise) to provide the most accurate picture. The optimization, however, utilizes the trained models.   

\begin{figure}[]
    \centering
    \includegraphics[scale=0.5]{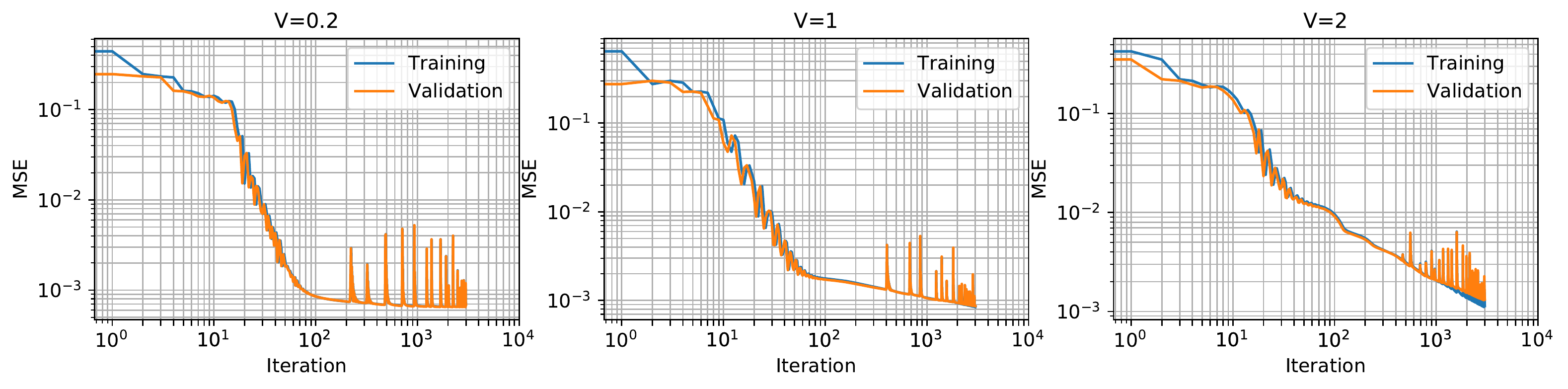}
    \caption{Evaluation of graybox model training and testing for multi-axis control applied to the system in the fermionic bath when varying the coupling strength. The number of shots is fixed at $N=1024$.}
    \label{fig:F_MA_V}
\end{figure}

\begin{figure}[]
    \centering
    \includegraphics[scale=0.5]{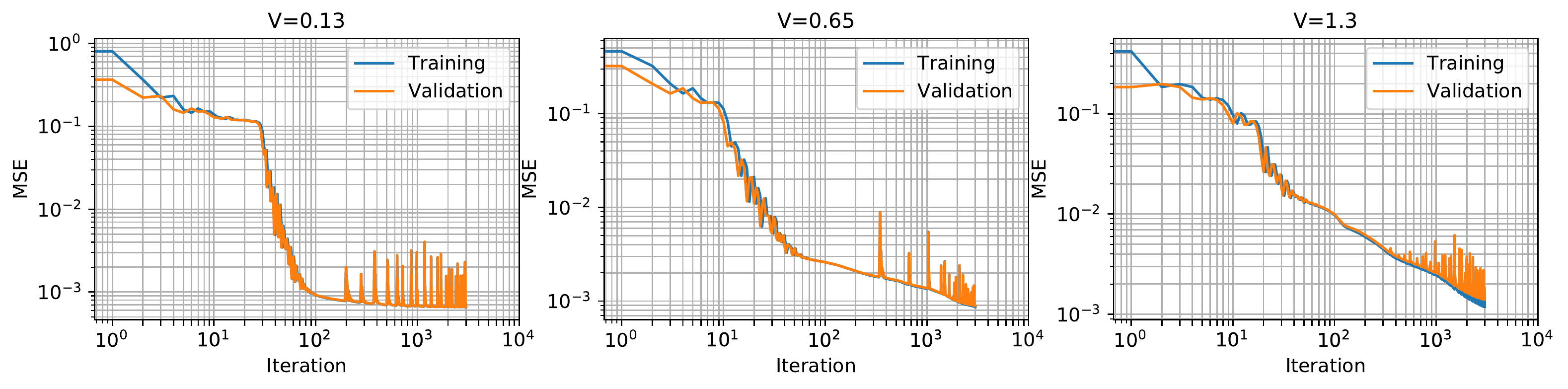}
    \caption{Evaluation of graybox model training and testing for multi-axis control applied to the system in the bosonic bath when varying the coupling strength. The number of shots is fixed at $N=1024$.}
    \label{fig:B_MA_V}
\end{figure}

For the second set of experiments, where we study the effect of the coupling strength, we see from Figs. \ref{fig:F_MA_V} and \ref{fig:B_MA_V}, that with increasing strength parameter, the model accuracy decreases. This behaviour is expected. When the coupling to the bath is very weak, (i.e. $V \to 0$), the system acts almost as a closed-system. In this case, the operator $V_O(T)$ is very close to the system identity operator $I$ almost independent of the control. In this case, the learning task is relatively easy, because the map between controls and $V_O(T)$ is almost a constant. When we increase the coupling strength, the map between the control and the $V_O(T)$ operator becomes non-trivial, and the model has to learn the dynamics of how control affects $V_O(T)$ indirectly from the data. Since, we are fixing the architecture of the ML model (particularly the number of hidden nodes in the GRU layers) for the comparison, we see the performance affected by increasing $V$. In other words, if we want to maintain the same model accuracy when increasing $V$, then we should increase the model complexity accordingly. In practice, $V$ is not known, and thus the ML model complexity is determined as a part of the design process where we try to optimize the hyperparameters to maximize model accuracy.

As for the control performance, from Table \ref{tab:compareV} we see that we can obtain high process fidelities for all gates, but with increasing coupling strength, the fidelities decrease. This can be attributed to the control constraints (maximum amplitude) relative to the noise strength. An interesting example is the Pauli Z-gate. While the multi-axis control clearly outperforms the single-axis in the first set of experiments where $V=2$ and $V=1.3$ for fermionic and bosonic baths, it is still significantly lower than other gates. In order to support the idea that this behaviour is related to the control constraints, we created another dataset with similar simulation parameters but chose the maximum allowed amplitude for the control pulses $A_{\rm max}=100$ instead of $A_{\rm max}=25$, and executed all the steps of the algorithm. At a coupling strength of $V=1$ and $V=0.65$ for the fermionic and bosonic baths respectively, and choosing the number of shots $N=1024$, we found that in the case of multi-axis control we get process fidelities of 96.36\% and 95.76\% for the two baths, which is higher than the ones reported in Table \ref{tab:compareV} of 95.82\% and 95.54\%, respectively. In conclusion, the stronger the noise is, the more difficult it is to control the system in the presence of  constraints. This demonstrates the importance of studying controllability of open quantum systems, which still remains an open area of research and is beyond the scope of this paper.

\begin{table}[]
\begin{tabular}{|c|ccc|ccc|}
\hline
\multirow{2}{*}{Gate} & \multicolumn{3}{c|}{Fermionic}                                    & \multicolumn{3}{c|}{Bosonic}                                          \\ \cline{2-7} 
                      & \multicolumn{1}{c|}{$V=0.2$} & \multicolumn{1}{c|}{$V=1$} & $V=2$ & \multicolumn{1}{c|}{$V=0.13$} & \multicolumn{1}{c|}{$V=0.65$} & $V=1.3$ \\ \hline
$I$&\multicolumn{1}{c|}{99.89}&\multicolumn{1}{c|}{98.63}&\multicolumn{1}{c|}{95.35}&\multicolumn{1}{c|}{99.87}&\multicolumn{1}{c|}{98.29}&\multicolumn{1}{c|}{92.35}\\ \hline
$X$&\multicolumn{1}{c|}{99.87}&\multicolumn{1}{c|}{98.13}&\multicolumn{1}{c|}{93.52}&\multicolumn{1}{c|}{99.84}&\multicolumn{1}{c|}{97.69}&\multicolumn{1}{c|}{91.70}\\ \hline
$Y$&\multicolumn{1}{c|}{99.86}&\multicolumn{1}{c|}{97.83}&\multicolumn{1}{c|}{93.82}&\multicolumn{1}{c|}{99.79}&\multicolumn{1}{c|}{97.55}&\multicolumn{1}{c|}{91.49}\\ \hline
$Z$&\multicolumn{1}{c|}{99.85}&\multicolumn{1}{c|}{95.82}&\multicolumn{1}{c|}{88.31}&\multicolumn{1}{c|}{99.79}&\multicolumn{1}{c|}{95.54}&\multicolumn{1}{c|}{85.71}\\ \hline
$H$&\multicolumn{1}{c|}{99.89}&\multicolumn{1}{c|}{96.95}&\multicolumn{1}{c|}{90.84}&\multicolumn{1}{c|}{99.82}&\multicolumn{1}{c|}{97.61}&\multicolumn{1}{c|}{91.45}\\ \hline
$pi_4$&\multicolumn{1}{c|}{99.90}&\multicolumn{1}{c|}{98.78}&\multicolumn{1}{c|}{95.52}&\multicolumn{1}{c|}{99.88}&\multicolumn{1}{c|}{98.47}&\multicolumn{1}{c|}{93.17}\\ \hline
\end{tabular}
\caption{The performance of the control in terms of process fidelity, when varying the coupling of the qubit to the auxiliary system, for both fermionic and bosonic baths. In this experiment, we fix the number of shots to $N=1024$, and show results for multi-axis control using gradient-descent method}
\label{tab:compareV}
\end{table}

\section{Conclusion}
\label{sec:conclu}
In this paper, we have presented a detailed numerical study for an ML approach of controlling open quantum systems with non-Markovian quantum noise. We showed the success of the method for the case when the underlying quantum noise comes from  fermionic or bosonic quantum baths. In order to facilitate the simulation of the bath to create synthetic datasets, we use an auxiliary system coupled to bosonic or fermionic white noise process as a model for the non-Markovian quantum noise and consider the reduced dynamics of the qubit. In an experimental setting, we only need standard prepare-control-measure procedures without the need of any prior information about the environment. We showed results for both single-axis and multi-axis control for a set of universal quantum gates. There are many interesting extensions to this work. This includes extending this approach to continuous-time measurements as well as continuous-variable systems, and applying and testing this method to an experimental hardware platform.

\paragraph*{Acknowledgments}
AY is funded by the Australian Government through the Australian Research Council under the Centre of Excellence Scheme No. CE170100012.
\bibliographystyle{apsrev4-1}
\bibliography{references} 
\end{document}